\documentclass[%
 aip,
 amsmath,amssymb,
preprint,%
]{revtex4-1}

\usepackage{graphicx}
\usepackage{epstopdf}

\usepackage{dcolumn}
\usepackage{bm}

\usepackage[utf8]{inputenc}
\usepackage[T1]{fontenc}
\usepackage{mathptmx}
\usepackage{etoolbox}

\makeatletter
\def\@email#1#2{%
 \endgroup
 \patchcmd{\titleblock@produce}
  {\frontmatter@RRAPformat}
  {\frontmatter@RRAPformat{\produce@RRAP{*#1\href{mailto:#2}{#2}}}\frontmatter@RRAPformat}
  {}{}
}%
\makeatother

\begin{document}

\title{Volume-preserving particle integrator based on exact flow of velocity
for nonrelativistic particle-in-cell simulations}
\author{Tsunehiko N. Kato}%
 \email[Corresponding author. E-mail address: ]{tsunehiko.kato@nao.ac.jp.}
\affiliation{Center for Computational Astrophysics, Notional Astronomical Observatory of Japan, 2-21-1 Osawa, Mitaka, Tokyo 181-8588, Japan}
\author{Seiji Zenitani}
\affiliation{Research Center for Urban Safety and Security, Kobe University, 1-1 Rokkodai-cho, Nada-ku, Kobe, 657-8501, Japan}

\date{\today}

\begin{abstract}
We construct a particle integrator for nonrelativistic particles
by means of the splitting method
based on the exact flow of the equation of motion
of particles in the presence of constant electric and magnetic field.
This integrator is volume-preserving similar to the standard Boris integrator
and is suitable for long-term integrations in particle-in-cell simulations.
Numerical tests reveal that
it is significantly more accurate than
previous volume-preserving integrators with second-order accuracy.
For example,
in the $E \times B$ drift test,
this integrator is more accurate than the Boris integrator
and the integrator based on the exact solution of gyro motion
by three and two orders of magnitude, respectively.
In addition,
we derive approximate integrators that incur low computational cost
and 
high-precision integrators displaying fourth- to tenth-order accuracy
with the aid of the composition method.
These integrators are also volume-preserving.
It is also demonstrated that
the Boris integrator is equivalent to
the simplest case of the approximate integrators derived in this study.
\end{abstract}

\maketitle

\section{Introduction}

The particle-in-cell (PIC) method\cite{Hockney1981, Birdsall1991} is 
a plasma simulation method
that can address the kinematic phenomena in plasmas.
It has been widely used to investigate various phenomena
in plasma physics, astrophysics, space physics, etc.
The method consists of two parts.
One part solves the motion of charged particles in an electromagnetic field,
and the other part solves the time evolution of the electromagnetic field
owing to charges and currents caused by the charged particles.
The solver in the former part is called particle integrator.

The most standard particle integrator is the Boris integrator\cite{Hockney1981, Birdsall1991, Boris1970}.
It is a simple second-order precision method.
Particle integrators with improved accuracy have been proposed in recent years.
These include methods that exactly solve the gyro motion part of the equation of motion
of particle (e.g., the $G_h^2$ method of \citet{He2015}
and Boris-C solver of \citet{ZenitaniUmeda2018})
and methods that improve the accuracy
while keeping calculation cost moderately low
by dividing the gyro motion step of the Boris integrator
into smaller substeps and calculating these\cite{Umeda2018, ZenitaniKato2020}.
Henceforth, we call the above two classes of integrators as ``exact gyration integrator''
and ``subcycled Boris integrator,'' respectively.
(Essentially,
the Boris integrator in the original form\cite{Boris1970} is the exact gyration integrator.
However,
what is widely used as the standard Boris method\cite{Hockney1981, Birdsall1991} at present
is an approximation of the original one.
In this paper also,
we call the approximated version as ``Boris integrator.''
See pg.~24 of Ref.~\onlinecite{Boris1970} and Ref.~\onlinecite{ZenitaniUmeda2018} for details.)

The volume-preserving property of integrators in phase space
is important for the long-term accuracy of integrations \cite{Qin2013}.
All the aforementioned integrators display this property.
Meanwhile,
the Runge--Kutta method (for example) with fourth-order accuracy
does not display this property.
Although it is more accurate than the Boris integrator at each step,
it deteriorates in long-term integrations\cite{Qin2013, He2015}
because errors accumulate with time.
Thus,
particle integrators are required to display this property.

A systematic method
to construct a particle integrator that satisfies the volume-preservation condition\cite{Zhang2015}
is provided with the aid of the splitting method\cite{McLachlan2002, Hairer2006}.
In the splitting method,
the vector field of the differential equation
(which prescribes the time development of the solution of the differential equation)
is decomposed into several sub-vector fields.
Then,
each sub-vector field defines a subsystem of differential equations
that can be solved more conveniently than the original equation.
Finally,
an approximate solution of the original differential equation is constructed
by combining exact or approximate solutions of
the respective differential equations of the subsystems
that satisfy the volume-preservation condition.
The recently developed volume-preserving integrators mentioned above
can be considered within this framework.
In such integrators,
the part of the vector field for the equation of motion
is decomposed further 
into a part involving an electric field
and another part involving a magnetic field.
Meanwhile,
a more accurate integrator is likely to be obtained
if the exact solution of the equation of motion
without such decomposition 
is used directly to construct the integrator.

In this study, 
we construct a volume-preserving particle integrator for nonrelativistic particles
by means of the splitting method
using the exact solution of the equation of motion
of particles in the presence of both constant electric and magnetic field.
In addition,
two classes of approximate integrators
that reduce the calculation cost are derived.
It is demonstrated that
the Boris integrator is equivalent to
the simplest case of one of the two methods.
We also derive higher-order integrators
with the aid of the composition method,
for highly accurate integrations.

\section{Exact velocity integrator}

\subsection{Splitting method}
The motion of a nonrelativistic particle with mass $m$ and charge $q$
in an electromagnetic field ($\bm{E}$ and $\bm{B}$)
is determined by the following differential equations:
\begin{equation}
\frac{d\bm{x}}{dt} = \bm{v},
\qquad
    \frac{d\bm{v}}{dt} = \frac{q}{m} \left( \bm{E} + \bm{v} \times \bm{B} \right).
\label{eq:eom0}
\end{equation}
By introducing an independent variable $s$ rather than $t$ such that
\begin{equation}
\frac{dt}{ds} = 1
\end{equation}
and defining a vector
\begin{equation}
y \equiv (t, \bm{x}, \bm{v}),
\end{equation}
the above system of differential equations (\ref{eq:eom0}) becomes
an autonomous one for $y$:
\begin{equation}
\frac{dy}{ds} = X(y).
\label{eq:y_dif_eq}
\end{equation}
The vector-valued function $X(y)$ on the right-hand-side of this equation
is called the vector field.
It prescribes the time development of the solution $y$.
For a time interval $h$,
a mapping between the solutions from $y(s)$ to $y(s+h)$
for an arbitrary $s$
is called the flow of the differential equation. It is denoted by $\varphi_h$:
\begin{equation}
\varphi_h:\quad
y(s) \to y(s + h).
\end{equation}
An instance of application of this mapping corresponds to one step of numerical integration of Eq.~(\ref{eq:y_dif_eq})
with the step size $h$.
However,
it is generally difficult to obtain the exact flow directly.

We adopt the splitting method \cite{McLachlan2002, Hairer2006} to obtain an approximated numerical integrator for Eq.~(\ref{eq:y_dif_eq}).
According to the steps presented in Ref.~\onlinecite{Zhang2015},
first,
we decompose the vector field $X(y)$
into the following sub-vector fields:
\begin{equation}
X(y) = X_t(y) + X_x(y) + X_v(y),
\end{equation}
where
\begin{equation}
X_t(y) = (1, 0, 0),
\qquad
X_x(y) = (0, \bm{v}, 0),
\qquad 
X_v(y) = \left(0, 0, \frac{q}{m} \left( \bm{E} + \bm{v} \times \bm{B} \right) \right).
\end{equation}
Then,
corresponding to these,
the equation (\ref{eq:y_dif_eq}) is split into the following subsystems:
\begin{equation}
X_t:\quad  \frac{dt}{ds} = 1,
\quad \frac{d\bm{x}}{ds} = 0,
\quad \frac{d\bm{v}}{ds} = 0
\end{equation}
\begin{equation}
X_x:\quad  \frac{dt}{ds} = 0,
\quad \frac{d\bm{x}}{ds} = \bm{v},
\quad \frac{d\bm{v}}{ds} = 0
\end{equation}
\begin{equation}
X_v:\quad  \frac{dt}{ds} = 0,
\quad \frac{d\bm{x}}{ds} = 0,
\quad \frac{d\bm{v}}{ds} = \frac{q}{m} \left( \bm{E} + \bm{v} \times \bm{B} \right).
\label{eq:Xv_eqs}
\end{equation}
If the flows of these subsystems are obtained by $\varphi^t_h, \varphi^x_h$, and $\varphi^v_h$,
respectively,
an approximation of the flow of the overall system (denoted by $\Phi_h$)
is constructed by a combination of these.
In particular, the composition
\begin{equation}
\Phi_h = \varphi^t_{h/2} \circ \varphi^x_{h/2} \circ \varphi^v_{h} \circ \varphi^x_{h/2} \circ \varphi^t_{h/2}
\label{eq:Strang_splitting}
\end{equation}
is a case of Strang splitting, which is symmetric (i.e., $\Phi_h=\Phi_{-h}^{-1}$) and
is a second-order approximation of the exact overall flow $\varphi_h$.
The exact flows of $\varphi^t_h$ and $\varphi^x_h$ can be obtained as
\begin{equation}
\varphi^t_h:\quad t \to t + h,
\qquad
\varphi^x_h:\quad \bm{x} \to \bm{x} + \bm{v} h.
\label{eq:t_x_flow}
\end{equation}
The exact flow $\varphi^v_h$ can also be obtained
as demonstrated in the following subsection.

\subsection{Exact flow of velocity for nonrelativistic charged particles
in a constant electromagnetic field}
\label{subsec:exact_flow}
According to the splitting method (\ref{eq:Xv_eqs}),
the flow for the velocity part, $\varphi^v_h$, is determined by the following equation of motion:
\begin{equation}
\frac{d\bm{v}}{ds} =  \frac{q}{m} \left( \bm{E} + \bm{v} \times \bm{B} \right),
\label{eq:eom}
\end{equation}
whereas the variables other than the velocity 
(i.e., time and position) 
are regarded as constants while considering this differential equation.
Therefore,
the electromagnetic fields are evaluated at fixed time and position,
and are also regarded as constants.

By defining the normalized electromagnetic fields
\begin{equation}
\tilde{\bm{E}} \equiv \frac{q}{m} \bm{E}
\qquad \textrm{and} \qquad
\tilde{\bm{B}} \equiv \frac{q}{m} \bm{B},
\end{equation}
the equation (\ref{eq:eom}) can be rewritten as
\begin{equation}
\frac{d\bm{v}}{ds} = \tilde{\bm{E}} + \bm{v} \times \tilde{\bm{B}}.
\label{eq:eom2}
\end{equation}
As shown in Appendix~\ref{sec:exact_solution_of_eom},
the exact solution of this equation is given by
\begin{equation}
\bm{v}(s) = \bm{v}(0)
+ f_1 \bm{e}_1
+ f_2 \bm{e}_2
+ f_3 \bm{e}_3.
\label{eq:v_solution_3}
\end{equation}
Here, we define the phase angle
\begin{equation}
\theta \equiv \tilde{B}s
\end{equation}
with $\tilde{B} \equiv |\tilde{\bm{B}}|$;
the following factors, which are functions of $\theta$,
\begin{equation}
f_1 \equiv \frac{\sin\theta}{\tilde{B}},
\qquad
f_2 \equiv \frac{1 - \cos\theta}{\tilde{B}^2},
\qquad
f_3 \equiv \frac{\theta - \sin\theta}{\tilde{B}^3};
\label{eq:f_def}
\end{equation}
and the ``bases''
\begin{equation}
\bm{e}_1 \equiv \tilde{\bm{E}} + \bm{v}(0) \times \tilde{\bm{B}},
\qquad
\bm{e}_2 \equiv \bm{e}_1 \times \tilde{\bm{B}},
\qquad
\bm{e}_3 \equiv (\tilde{\bm{E}}\cdot \tilde{\bm{B}}) \tilde{\bm{B}}.
\end{equation}
The factors $f_1$, $f_2$, and $f_3$ converge for $\tilde{B}, \theta \to 0$ as
\begin{equation}
f_1 \to s,
\qquad
f_2 \to \frac{1}{2} s^2,
\qquad
f_3 \to \frac{1}{6} s^3.
\end{equation}
When $\tilde{B}$ and $\theta$ are highly marginal,
these factors can be evaluated numerically using the Taylor expansions of
the sine and cosine functions.

In Eq.~(\ref{eq:v_solution_3}),
we can arbitrarily fix the value of the initial velocity $\bm{v}(0)$.
Therefore,
the exact flow $\varphi^v_h$ is obtained from Eq.~(\ref{eq:v_solution_3})
as a mapping for any initial velocity $\bm{v}$ as follows:
\begin{equation}
\varphi_h^v:
\quad
\bm{v} \to \bm{v} + f_1\bm{e}_1 + f_2\bm{e}_2 + f_3\bm{e}_3,
\label{eq:v_flow}
\end{equation}
with $\theta = \tilde{B}h$.

\subsection{Volume-preservation condition}
The volume-preserving property of the flow is important
for the accuracy of long-term numerical integrations.
For a flow $\varphi$,
the condition for volume-preservation is given by
the Jacobian determinant as
\begin{equation}
\left|\frac{\partial\varphi(y)}{\partial y}\right|  = 1
\qquad \textrm{for any $y$}.
\end{equation}
It is established that
exact flows of divergence-free vector-field satisfy this condition
(Liouville's theorem)\cite{Zhang2015}.
Here, a vector-field is divergence-free if
\begin{equation}
\nabla_y \cdot X = 0.
\end{equation}
It can be conveniently shown that
the vector-fields for the original system $X$
as well as those for the decomposed subsystems $X_t$, $X_x$, and $X_v$
are divergence-free.
Hence, the following hold:
\begin{equation}
\left|\frac{\partial\varphi_h(t)}{\partial t}\right| = 1,
\qquad
\left|\frac{\partial\varphi^t_h(t)}{\partial t}\right| = 1,
\qquad
\left|\frac{\partial\varphi^x_h(\bm{x})}{\partial \bm{x}}\right| = 1,
\qquad
\left|\frac{\partial\varphi^v_h(\bm{v})}{\partial \bm{v}}\right| = 1.
\label{eq:volume_preserving_condition}
\end{equation}
For the approximated overall flow $\Phi_h(y)$,
the volume-preservation condition is given by
\begin{equation}
\left|\frac{\partial\Phi_h(y)}{\partial y}\right|  = 1
\qquad \textrm{for any $y$}.
\label{eq:volume_preserving_condition_Phi}
\end{equation}
Because this flow is constructed using the exact flows of the subsystem according to Strang splitting (\ref{eq:Strang_splitting})
and these satisfy the volume-preservation condition (\ref{eq:volume_preserving_condition}),
the condition (\ref{eq:volume_preserving_condition_Phi}) is also satisfied
and the flow $\Phi_h$ is volume-preserving.

For subsequent convenience,
we derive an explicit expression of
the volume-preservation condition for the flow $\varphi^v_h$.
In this case,
the Jacobian matrix  and its determinant
are given by
\begin{equation}
\frac{\partial \varphi^v_h(\bm{v})}{\partial \bm{v}} =
\begin{pmatrix}
1 - f_2(\tilde{B}_y^2 +\tilde{B}_z^2) &
f_1\tilde{B}_z +f_2\tilde{B_x}\tilde{B}_y &
-f_1 \tilde{B}_y + f_2 \tilde{B}_x \tilde{B}_z\\
-f_1 \tilde{B}_z + f_2 \tilde{B}_x \tilde{B}_y &
1 - f_2(\tilde{B}_x^2 + \tilde{B}_z^2) &
f_1 \tilde{B}_x + f_2\tilde{B}_y\tilde{B}_z\\
f_1 \tilde{B}_y + f_2 \tilde{B}_x \tilde{B_z} &
-f_1 \tilde{B}_x + f_2 \tilde{B}_y \tilde{B}_z &
1 - f_2 (\tilde{B}_x^2 + \tilde{B}_y^2)
\end{pmatrix},
\end{equation}
and
\begin{equation}
\left| \frac{\partial \varphi^v_h(\bm{v})}{\partial \bm{v}} \right| =
f_1^2 \tilde{B}^2 + \left(1 - f_2 \tilde{B}^2\right)^2.
\label{eq:phi_v_Jacobi}
\end{equation}
By substituting the definitions of $f_1$ and $f_2$ in Eq.~(\ref{eq:f_def}) into this expression,
we can verify that the volume-preservation condition in Eqs.~(\ref{eq:volume_preserving_condition})
is satisfied.
The equation (\ref{eq:phi_v_Jacobi}) with
the condition in Eq.~(\ref{eq:volume_preserving_condition}) would be used subsequently
while deriving the approximate methods.

\subsection{Exact velocity integrator}
The above results can be used to construct a particle integrator for the PIC simulations.
Let the time step of the simulation be $h = \Delta t$.
According to Strang splitting (\ref{eq:Strang_splitting}),
a simulation step from Step $n$ to Step $n+1$ is split into the following substeps:
\begin{align*}
&1.\ \varphi^t_{\Delta t/2}:  \quad t^n \to t^{n+1/2} = t^n + \frac{1}{2}\Delta t\\
&2.\ \varphi^x_{\Delta t/2}: \quad \bm{x}^n \to \bm{x}^{n+1/2} = \bm{x}^n + \frac{1}{2}\bm{v}^n \Delta t\\
&3.\ \varphi^v_{\Delta t}:  \quad \bm{v}^n \to \bm{v}^{n+1}\\
&4.\ \varphi^x_{\Delta t/2}:  \quad \bm{x}^{n+1/2} \to \bm{x}^{n+1} = \bm{x}^{n+1/2} + \frac{1}{2}\bm{v}^{n+1} \Delta t\\
&5.\ \varphi^t_{\Delta t/2}:  \quad t^{n+1/2} \to t^{n+1} = t^{n+1/2} + \frac{1}{2}\Delta t
\end{align*}
The update of the velocity in Substep 3 is given by Eq.~(\ref{eq:v_flow}).
For subsequent convenience, we rewrite it explicitly as follows:
\begin{equation}
\bm{v}^{n+1} = \bm{v}^{n}
+ f_1 \bm{e}_1
+ f_2 \bm{e}_2
+ f_3 \bm{e}_3,
\label{eq:EV_integrator}
\end{equation}
where
\begin{equation}
\theta = \tilde{B} \Delta t,
\qquad
f_1 = \frac{\sin\theta}{\tilde{B}},
\qquad
f_2 = \frac{1 - \cos\theta}{\tilde{B}^2},
\qquad
f_3 = \frac{\theta - \sin\theta}{\tilde{B}^3}
\label{eq:EV_f}
\end{equation}
and
\begin{equation}
\bm{e}_1 \equiv \tilde{\bm{E}} + \bm{v}^{n} \times \tilde{\bm{B}},
\qquad
\bm{e}_2 \equiv \bm{e}_1 \times \tilde{\bm{B}},
\qquad
\bm{e}_3 \equiv (\tilde{\bm{E}}\cdot \tilde{\bm{B}}) \tilde{\bm{B}}.
\end{equation}
Because of Substeps 1 and 2 above,
the electric and magnetic fields should be evaluated at
time $t^{n+1/2}$ and position $\bm{x}^{n + 1/2}$:
\begin{equation}
\tilde{\bm{E}}=\frac{q}{m}\bm{E}(t^{n+1/2}, \bm{x}^{n+1/2}),
\qquad
\tilde{\bm{B}}=\frac{q}{m}\bm{B}(t^{n+1/2}, \bm{x}^{n+1/2}).
\end{equation}
Considering the numerical accuracy,
it would be more effective to evaluate
the factor $1-\cos(\theta)$ (which appears in the calculation of $f_2$)
using the relationship
\begin{equation}
1 - \cos(\theta) = 2 \sin^2(\theta/2).
\end{equation}
As mentioned in Subsection \ref{subsec:exact_flow},
when $\tilde{B}$ and $\theta$ are highly marginal,
the factors $f_i$ can be evaluated using the Taylor expansions of
the sine and cosine functions.
For example,
the following expressions are obtained by truncating the series up to the order $\theta^2$.
\begin{equation}
f_1 = \Delta t,
\qquad
f_2 = \frac{1}{2} \Delta t^2,
\qquad
f_3 = 0.
\end{equation}
We define a threshold $\theta_c$
that is marginal enough for the above expressions to be effective.
Thus, the factors can be evaluated using these expressions
when $|\theta| < \theta_c$.
This procedure defines a particle integrator,
and we call it as ``exact velocity integrator.''
As shown in the previous subsection,
it satisfies the volume-preservation condition.
Note that in this method, although all the decomposed flows are exact,
the overall accuracy of the integrator is second-order
owing to Strang splitting.

\subsection{Exact position--velocity integrator}
The exact solution of particle position is obtained as follows by integrating Eq.~(\ref{eq:v_solution_3}) over $s$:
\begin{equation}
\bm{x}(s) = \bm{x}(0) + \bm{v}(0)s
+ f_2 \bm{e}_1
+ f_3 \bm{e}_2
+ \frac{\frac{1}{2}s^2 - f_2}{\tilde{B}^2} \bm{e}_3.
\label{eq:x_solution_3}
\end{equation}
It is feasible to use this solution to construct another integrator.
However,
this method is not consistent with Strang splitting (\ref{eq:Strang_splitting}),
and the position for evaluating electromagnetic fields should be
provided independently.
With the electromagnetic fields determined in this manner,
the update of the particle position is given by the exact solution (\ref{eq:x_solution_3}) as
\begin{equation}
\bm{x}^{n+1} = \bm{x}^{n} + \bm{v}^{n}\Delta t
+ f_2 \bm{e}_1
+ f_3 \bm{e}_2
+ \frac{\frac{1}{2}\Delta t^2 - f_2}{\tilde{B}^2} \bm{e}_3.
\end{equation}

When the position selected for evaluating the electromagnetic field, $\tilde{\bm{x}}$, is
identical to that in Strang splitting (\ref{eq:Strang_splitting}) by
\begin{equation}
\tilde{\bm{x}} = \bm{x}^n + \frac{1}{2} \bm{v}^{n}  \Delta t,
\end{equation}
we obtain a particle integrator.
Here, we call it as ``exact position-velocity integrator.''
This integrator provides the exact solutions of both position and velocity
when the electromagnetic field is constant.
In other cases, it would be highly accurate for a short time interval.
However,
it is not suitable for long-term integrations
(as would be demonstrated subsequently).
This is because it does not satisfy the volume-preservation condition
and therefore, is deficient in long-term accuracy.

\section{Approximate methods and Higher-order methods}
From the perspective of calculation cost,
in certain scenarios,
it may not be reasonable to evaluate the sine and cosine
in $f_1$, $f_2$, and $f_3$ in Eq.~(\ref{eq:EV_f}).
To address this, in this section,
we derive approximate integrators
by replacing $\sin(\theta)$ and $\cos(\theta)$
with certain approximated functions with low calculation cost ($\tilde{S}(\theta)$ and $\tilde{C}(\theta)$),
so that
\begin{equation}
f_1 = \frac{\tilde{S}(\theta)}{\tilde{B}},
\qquad
f_2 = \frac{1 - \tilde{C}(\theta)}{\tilde{B}^2},
\qquad
f_3 = \frac{\theta - \tilde{S}(\theta)}{\tilde{B}^3}.
\end{equation}
The time development of the velocity is calculated with
these approximated factors $f_1$, $f_2$, and $f_3$ and the expression (\ref{eq:EV_integrator}).
This method defines a numerical flow of the velocity $\Phi^v_h(\bm{v})$.
Its Jacobian determinant is also given in the form of Eq.~(\ref{eq:phi_v_Jacobi})
by replacing $\varphi^v_h$ with $\Phi^v_h$.
In this case,
the volume-preservation condition for the velocity flow becomes
\begin{equation}
\left| \frac{\partial \Phi^v_h(\bm{v})}{\partial \bm{v}} \right| =
\tilde{S}(\theta)^2 + \tilde{C}(\theta)^2 = 1.
\label{eq:phi_v_Jacobi_approx}
\end{equation}
Here,
we derive two classes of approximate integrators that satisfy this condition.

\subsection{$S_n$-method}
As a simple approximation,
we propose a method in which $\tilde{S}(\theta)$
is given by a truncated Taylor series of the sine function up to a given order $n$ (denoted by $S_n(\theta)$)
for $\theta \le \pi/2$. The corresponding $\tilde{C}(\theta)$ function
is determined so that it satisfies the volume-preservation condition (\ref{eq:phi_v_Jacobi_approx})
as follows:
\begin{equation}
\tilde{S}(\theta) = S_n(\theta),
\qquad
\tilde{C}(\theta) = \sqrt{1 - \tilde{S}(\theta)^2}.
\end{equation}
For the first several orders of $n$,
$S_n(\theta)$ are given as follows:
\begin{equation}
S_1(\theta) = \theta,
\quad
S_3(\theta) = \theta - \frac{\theta^3}{3!},
\quad
S_5(\theta) = \theta - \frac{\theta^3}{3!} + \frac{\theta^5}{5!}.
\end{equation}
Because the sine is an odd function,
the order of the series, $n$, is odd. Consequently, the order of the approximation accuracy becomes $n+1$.
For $\theta > \pi/2$,
although $\theta$ generally does not adopt such a large value in general simulations,
$\tilde{S}(\theta)$ and $\tilde{C}(\theta)$ are determined by $S_n(\theta)$ for $\theta \le \pi/2$ as follows:
\begin{equation}
\tilde{S}(\theta) = S_n(\pi - \theta),
\qquad
\tilde{C}(\theta) = -\sqrt{1 - \tilde{S}(\theta)^2}.
\end{equation}
This method provides a class of approximated integrators and is henceforth called ``$S_n$-method.''
The $S_n(\theta)$ function must satisfy the condition $|S_n(\theta)| \le 1$
so that the $\tilde{C}(\theta)$ function is real.
Therefore,
in certain cases, there is a maximum value that $\theta$ can adopt.
The maximum values of $\theta$ for the $S_1$-, $S_5$-, and $S_9$-methods
are $1$, $1.49132$, and $1.56816$, respectively.
This is not the case for the $S_3$- and $S_7$-methods.
Note that
regardless of the accuracy of the velocity flow,
the overall accuracy is of the second order owing to Strang splitting.
In general,
for approximations by the truncated Taylor series to be effective,
the argument $\theta$ must be small enough ($\theta \ll 1$).
Considering numerical accuracy,
it would be more effective to evaluate
the factor $1-\tilde{C}(\theta)$ (which appears in the calculation of $f_2$)
using the relationship
\begin{equation}
1 - \tilde{C}(\theta) = \frac{\tilde{S}(\theta)^2}{1 + \tilde{C}(\theta)}.
\end{equation}

Similarly,
it is feasible to construct integrators in which
$\tilde{C}(\theta)$ is given by a truncated Taylor series of the cosine function
up to a given order $n$, $C_n(\theta)$. 
The corresponding $\tilde{S}(\theta)$ function is determined to satisfy the volume-preservation condition (\ref{eq:phi_v_Jacobi_approx}).
However,
the numerical accuracy of
the Taylor series of the cosine function is lower when $\theta \ll 1$. Thereby,
this method would be at most as effective as the $S_n$-method.
Therefore, we do not consider this class of approximations henceforth.

\subsection{$T_n$-method}
Another approximation can be adopted
using the Taylor series of $\tan(\theta/2)$.
By truncating it up to a specified order $n$, $T_n(\theta/2)$,
the functions $\tilde{S}(\theta)$ and $\tilde{C}(\theta)$
are expressed by the following relationships similar to that for the trigonometric functions:
\begin{equation}
\tilde{S}(\theta) = \frac{2 T_n(\theta/2)}
{1 + T_n(\theta/2)^2},
\qquad
\tilde{C}(\theta) = \frac{1 - T_n(\theta/2)^2}
{1 + T_n(\theta/2)^2}.
\end{equation}
This method also satisfies the volume-preservation condition (\ref{eq:phi_v_Jacobi_approx}).
Henceforth,
this method is called ``$T_n$-method.''
For the first several orders of $n$,
$T_n(\theta)$ are given as follows:
\begin{equation}
T_1\left(\frac{\theta}{2}\right) = \frac{\theta}{2},
\quad
T_3\left(\frac{\theta}{2}\right) = \frac{\theta}{2} + \frac{1}{3} \left( \frac{\theta}{2} \right)^3,
\quad
T_5\left(\frac{\theta}{2}\right) = \frac{\theta}{2} + \frac{1}{3} \left( \frac{\theta}{2} \right)^3
+ \frac{2}{15} \left( \frac{\theta}{2} \right)^5.
\end{equation}
Similar to the $S_n$-method,
because the tangent is an odd function,
the order of the series, $n$, is odd, and the approximation order is $n+1$.
The overall accuracy of this method is of the second order.
This method has certain advantages compared with the $S_n$-method:
it can circumvent the need to evaluate the square root function,
and there is no limit to the value of $\theta$.
Considering numerical accuracy,
it would be more effective to evaluate
the factor $1-\tilde{C}(\theta)$ in the calculation of $f_2$
using the following relationship:
\begin{equation}
1 - \tilde{C}(\theta) = \tilde{S}(\theta) T_n(\theta/2).
\end{equation}
It should be noted that
the Boris integrator is equivalent to the simplest case of this method.
It can be demonstrated that the update of the velocity by the Boris integrator
is equivalent to the $T_1$-method.

\subsection{Higher-order methods}
\label{subsec:composition_method}
In certain cases,
numerical accuracy would be preferred over calculation speed.
Higher-order methods can be obtained conveniently using the composition method
with the aforementioned second-order methods
as basic components.
The triple jump method\cite{Yoshida1990} and Suzuki's fractal composition method\cite{Suzuki1990} are
similar composition methods.
In the present case,
both the methods provide fourth-order accuracy
because the basic second-order method, $\Phi_h$, is symmetric (time-reversible)\cite{Hairer2006}.
In the triple jump method,
the fourth-order flow $\Psi^{3J}_h$ is obtained
by the composition of three basic flows with different step sizes as follows:
\begin{equation}
\Psi^{3J}_h = \Phi_{\gamma_3 h} \circ \Phi_{\gamma_2 h} \circ \Phi_{\gamma_1 h},
\end{equation}
where the factors $\gamma_1$, $\gamma_2$ and $\gamma_3$ are given by
\begin{equation}
\gamma_1 = \gamma_3 =\frac{1}{2 - 2^{1/3}},
\qquad
\gamma_2 = - \frac{2^{1/3}}{2 - 2^{1/3}}.
\end{equation}
Similarly,
in Suzuki's fractal composition method,
the flow  $\Psi^{SZ}_h$
is given by
the composition of five basic flows:
\begin{equation}
\Psi^{SZ}_h = \Phi_{\gamma_5 h} \circ \Phi_{\gamma_4 h} \circ \Phi_{\gamma_3 h}
\circ \Phi_{\gamma_2 h} \circ \Phi_{\gamma_1 h},
\end{equation}
where
\begin{equation}
\gamma_1 = \gamma_2 = \gamma_4= \gamma_5 =\frac{1}{4 - 4^{1/3}},
\qquad
\gamma_3 = - \frac{4^{1/3}}{4 - 4^{1/3}}.
\end{equation}
Although this method is more calculation-intensive than the triple jump method,
it is generally more accurate.

Even a higher-order method can be constructed by the symmetric composition method\cite{Hairer2006}
by combining the basic flows as
\begin{equation}
\Psi_h = \Phi_{\gamma_n h} \circ \Phi_{\gamma_{n-1} h} \circ \cdots
\circ \Phi_{\gamma_2 h} \circ \Phi_{\gamma_1 h}.
\end{equation}
In the following, we use sixth-, eighth- and tenth-order methods.
The values of the factors $\gamma_i$ are given in Appendix \ref{sec:factor_symm_comp}.
The number of basic flows to be combined in the composition, $n$, are
7, 15, and 35, respectively.
Thus,
the calculation cost increases approximately twofold each time the order of accuracy increases by two.

\subsection{Compensated summation}
\label{subsec:compensated_summation}
For long-term integrations,
rounding errors generally accumulate
because of the many integration steps involved.
These errors can be problematic in high precision integrations.
Compensated summation\cite{Kahan1965}
is a technique for reducing the accumulation of rounding errors
in floating point arithmetic.
When the machine epsilon of the floating point arithmetic is $\varepsilon$,
this technique can effectively
calculate the summation similar to a scenario where the machine epsilon is $\varepsilon^2$.
For the summation of the following form 
\begin{equation}
y_{n+1} = y_n + \delta_n,
\end{equation}
where $\delta_n$ is the increment in the $n$-th step,
the compensated summation
is performed by the following procedure (see Section VIII.~5 in Ref.~\onlinecite{Hairer2006}):
\begin{align*}
&a = y_n\\
&e = e + \delta_n\\
&y_{n+1} = a + e\\
&e = e+(a - y_{n+1}),
\end{align*}
where the variable $e$ is initially set to zero and must be retained during
the series of the integration steps.
This technique should be applied when highly accurate calculations are required.
In the next section,
we apply it to numerical tests with the higher-order integrators.

\section{Numerical tests}

We have carried out several numerical tests to evaluate the accuracies of the new integrators
developed above.
In this section,
the results of the test simulations are shown.
In the following,
we consider the case $m = q = 1$ for the particle.

\subsection{$E\times B$ drift test}
\label{subsec:ExB_drift_test}
The results of test simulations to observe
the $E \times B$ drift of a particle in a constant electromagnetic field
are presented below.
The simulation parameters are
$\bm{E}=(0, 0.2, 0)$,
$\bm{B} = (0,0,1)$,
$\Delta t = 0.5$, $\bm{x}_0 = (0,0,0)$, and $\bm{v}_{0} = (1,0,0)$.
Here, $\bm{x}_0$ and $\bm{v}_0$ are the initial position and velocity,
respectively, of the particle.
In this case,
the $E \times B$ drift velocity is $V_D = 0.2$ in the $x$-direction,
the gyro period of the particle is $t_g = 2\pi$,
and the gyro radius is $r_g = 1$.
The angle $\theta$ becomes $\theta = \tilde{B}\Delta t = 0.5$,
which is approximately one-twelfth of $2\pi$ and relatively large 
for PIC simulations.
The calculation time is $ T = 2000 $,
which corresponds to approximately 320 gyro periods.

\begin{figure}
\centering
\includegraphics[width=16cm,clip]{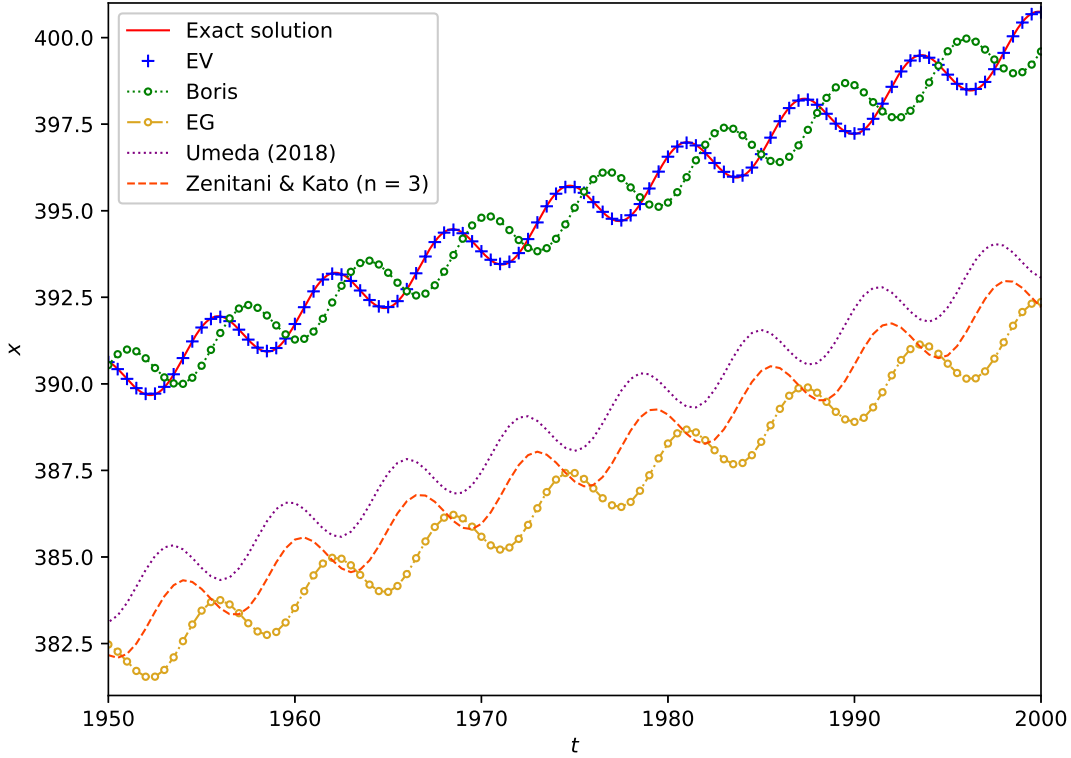}
\caption{
$x$-coordinate of particle versus time
near the completion of $E \times B$ drift simulation.
Results for the exact velocity integrator (EV),
the Boris integrator,
the exact gyration integrator (EG),
the subcycled Boris integrator by Umeda (2018),
and that by Zenitani and Kato (2020) of the subcycle number $n=3$
are shown.
The exact solution (\ref{eq:x_solution_3})
is also shown as a reference.
}
\label{fig:ExB_t_x}
\end{figure}

Figure \ref{fig:ExB_t_x} shows
the $x$-coordinate of the particle versus time
near the completion of the simulation.
The exact solution given by Eq.~(\ref{eq:x_solution_3}) is also shown as a reference.
Notwithstanding the consideration of a fairly large time-step,
the exact velocity integrator is almost in complete agreement with
the exact solution even near the 320 gyro periods.
(Although not presented here,
the exact position--velocity integrator is in complete agreement with
the exact solution because the electromagnetic field is constant in this case.)
For reference, the figure also plots the results obtained by the Boris integrator
for an equal time-step.
It can be observed that the motion of the gyrocenter is reproduced effectively.
However, the phase error is large.
The trajectories deviate substantially for the methods in which the accuracy of only the gyro motion part of the equation of motion
is improved
(such as the exact gyration integrator\cite{He2015, ZenitaniUmeda2018}
and the subcycled Boris integrators \cite{Umeda2018, ZenitaniKato2020}).

\begin{figure}
\includegraphics[width=16cm,clip]{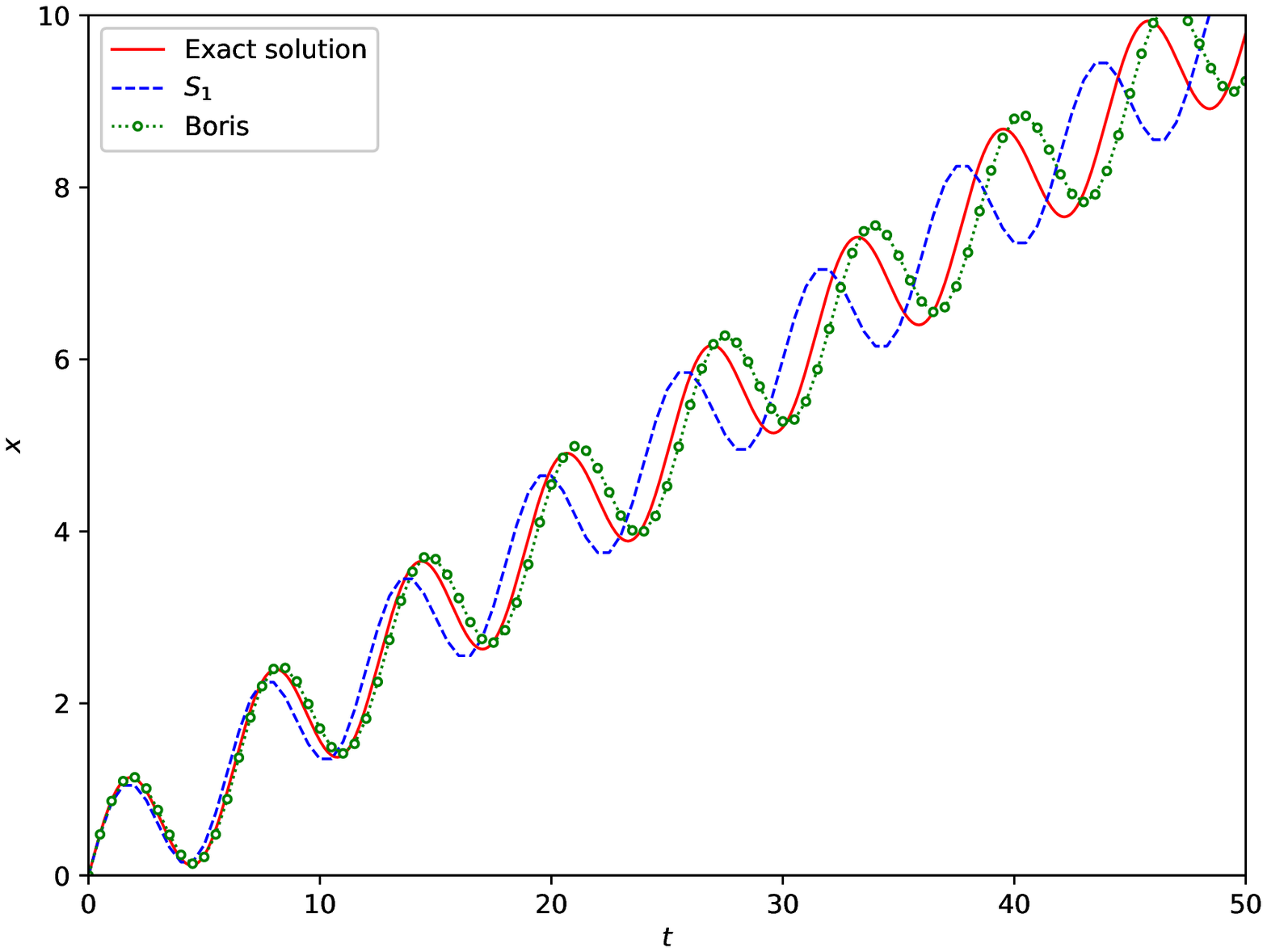}
\caption{
$x$-coordinate of particle versus time
in the early part of simulation.
The results of the $S_1$-method and Boris integrator ($T_1$-method)
are shown in conjunction with the exact solution.
}
\label{fig:ExB_t_x_2nd_order}
\end{figure}

Figure \ref{fig:ExB_t_x_2nd_order} shows the results obtained by
the methods with second-order accuracy of velocity flow,
the $S_1$-method, and the Boris integrator
(which is equivalent to the $T_1$-method).
It shows the early part of the simulation.
It is evident that both the integrators
are beginning to be out of phase.

\begin{figure}
\includegraphics[width=16cm,clip]{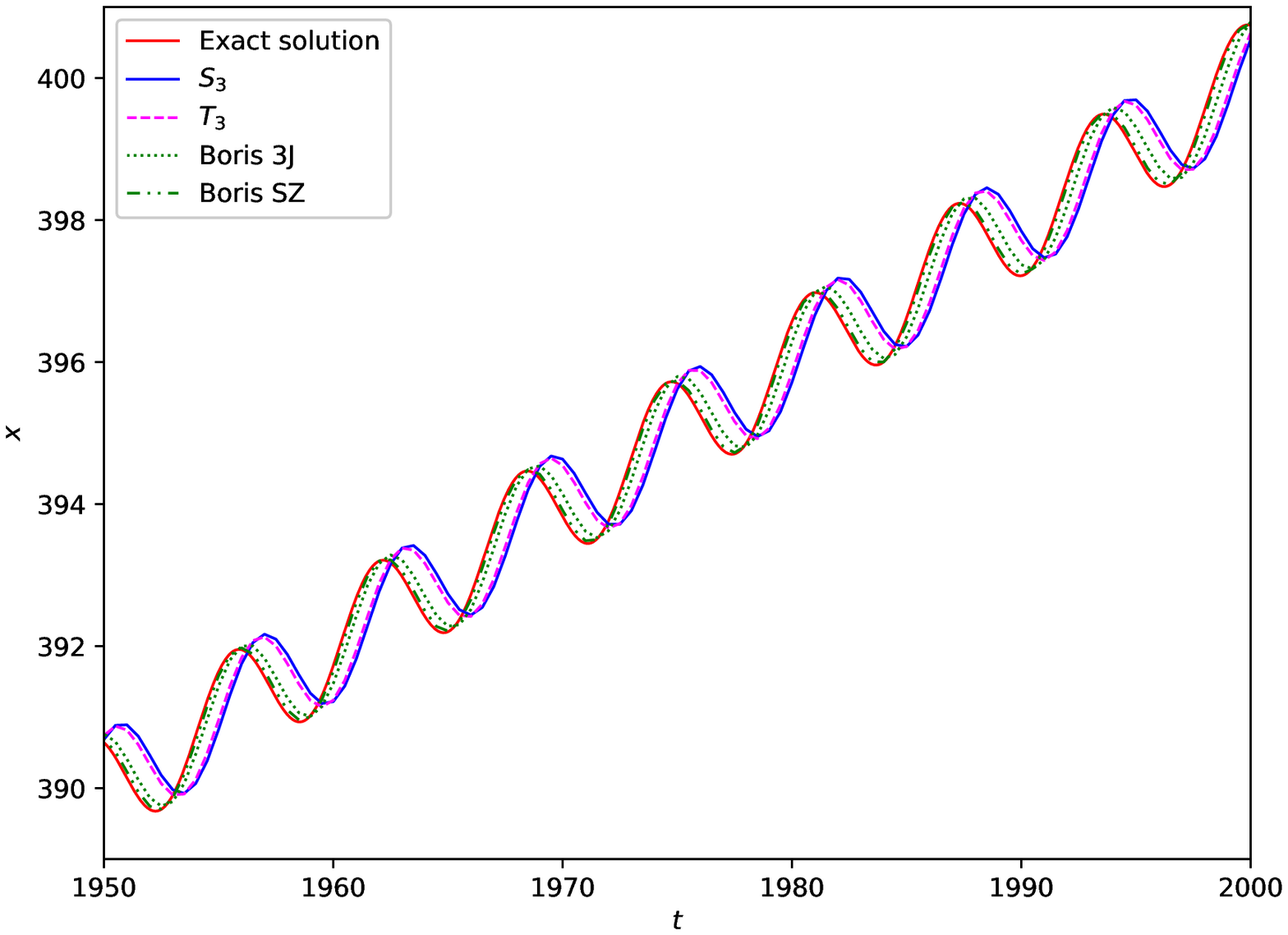}
\caption{
Similar to Fig.~\ref{fig:ExB_t_x}.
Results of the $S_3$-method, $T_3$-method,
the triple jump method applied to the Boris integrator (Boris 3J),
and Suzuki's fractal composition method applied to the Boris integrator (Boris SZ)
are shown with the exact solution.
}
\label{fig:ExB_t_x_4th_order}
\end{figure}

Figure~\ref{fig:ExB_t_x_4th_order} shows
the results of the methods with fourth-order accuracy of velocity flow,
$S_3$-method, and $T_3$-method.
Similar to Fig.~\ref{fig:ExB_t_x},
the results near the completion of the simulation are shown.
For reference,
this figure also shows the results obtained with
each of triple jump method and Suzuki's fractal composition method
applied to the Boris integrator as the basic method
(both display fourth-order accuracy in the overall flow).
In all the cases, the accuracy is improved substantially compared
with the methods in Fig.~\ref{fig:ExB_t_x_2nd_order}.
However, a marginal phase error also occurs.

\begin{figure}
\includegraphics[width=16cm,clip]{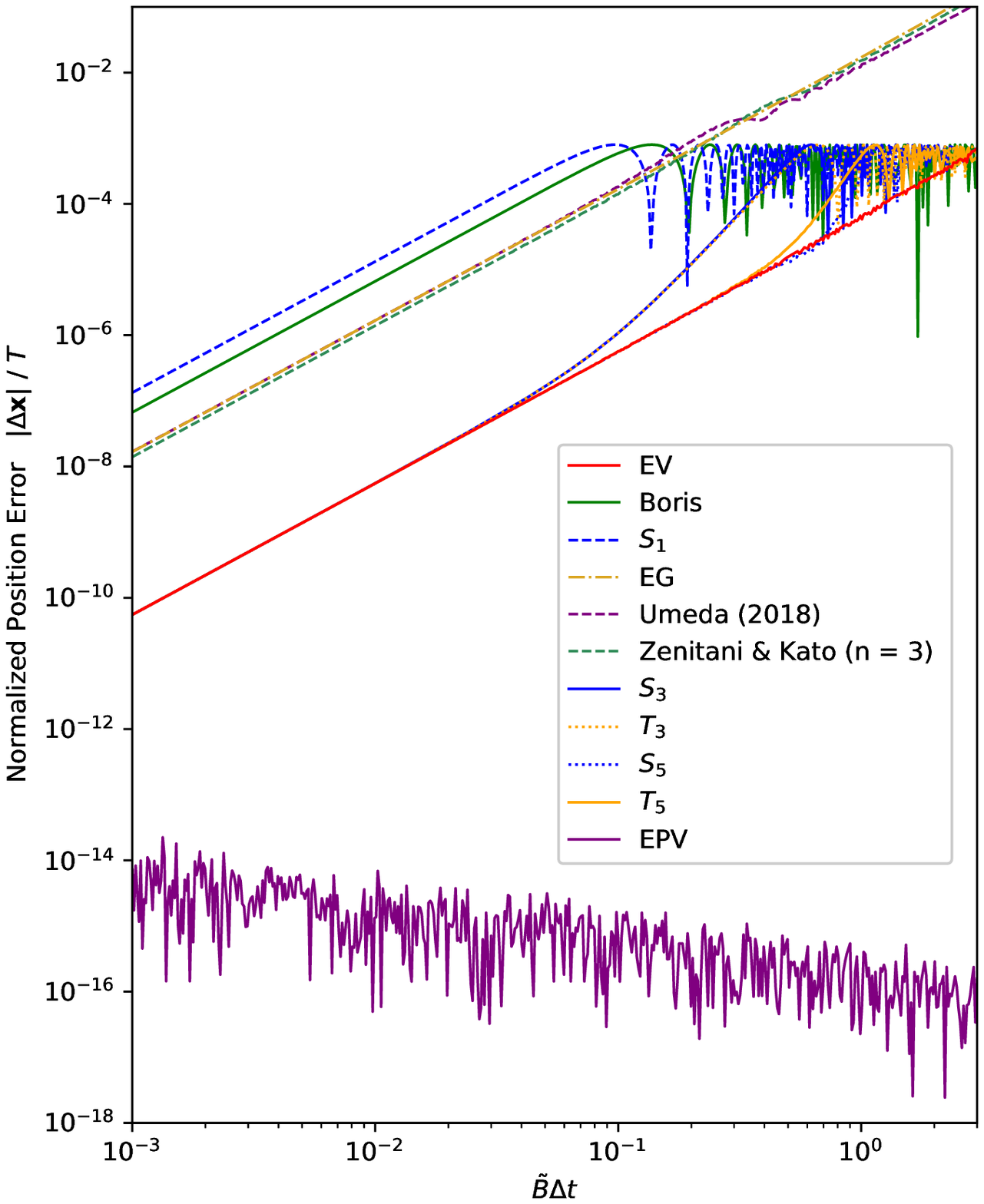}
\caption{
Magnitude of
error in particle position with respect to exact solution
per unit time,
$|\Delta \bm{x}| / T$, as a function of $\theta = \tilde{B}\Delta t$.
Results of the exact velocity integrator (EV),
methods with second- and fourth-order accuracy of the velocity flow,
and exact position-velocity integrator (EPV)
are shown.
}
\label{fig:ExB_pos_err}
\end{figure}

Figure~\ref{fig:ExB_pos_err} plots the magnitude of
the error in the particle position per unit time in the various methods,
$|\Delta \bm{x}| / T$, as functions of time.
Here, $T=2000$ is the time for evaluation, and
$|\Delta \bm{x}| $ is the magnitude of the position error
with respect to the exact solution at that time.
The error of the exact velocity integrator
is smaller than those of the Boris integrator
and exact gyration integrator by approximately three and two orders, respectively.
The results of the $S_3$-method and $T_3$-method
(which display fourth-order accuracy in the velocity flow)
almost coincide with that of the exact velocity integrator in the region where $\Delta t$ is small.
However, the error increases where $\Delta t$ is large.
Nonetheless, it is significantly more accurate than the Boris integrator.
The exact gyration integrator and subcycled Boris integrators
are more accurate than the Boris integrator when $\Delta t$ is small.
However,
when $\Delta t$ is large (see Fig. 1),
the trajectories deviate substantially, 
and the position error is larger than that of the Boris integrator.
Note that the results of the $S_n$-method and $T_n$-method are shown only 
for $n = 1, 3$, and $5$ in this figure
(as mentioned earlier, $T_1$-method is equivalent to the Boris integrator).
This is because when $n \ge 7$,
the position error is almost equal to that of the exact velocity integrator.
The exact position-velocity integrator solves
the trajectory accurately with the maximum precision,
except for the accumulation of rounding errors. This is mentioned subsequently.

\begin{figure}
\includegraphics[width=16cm,clip]{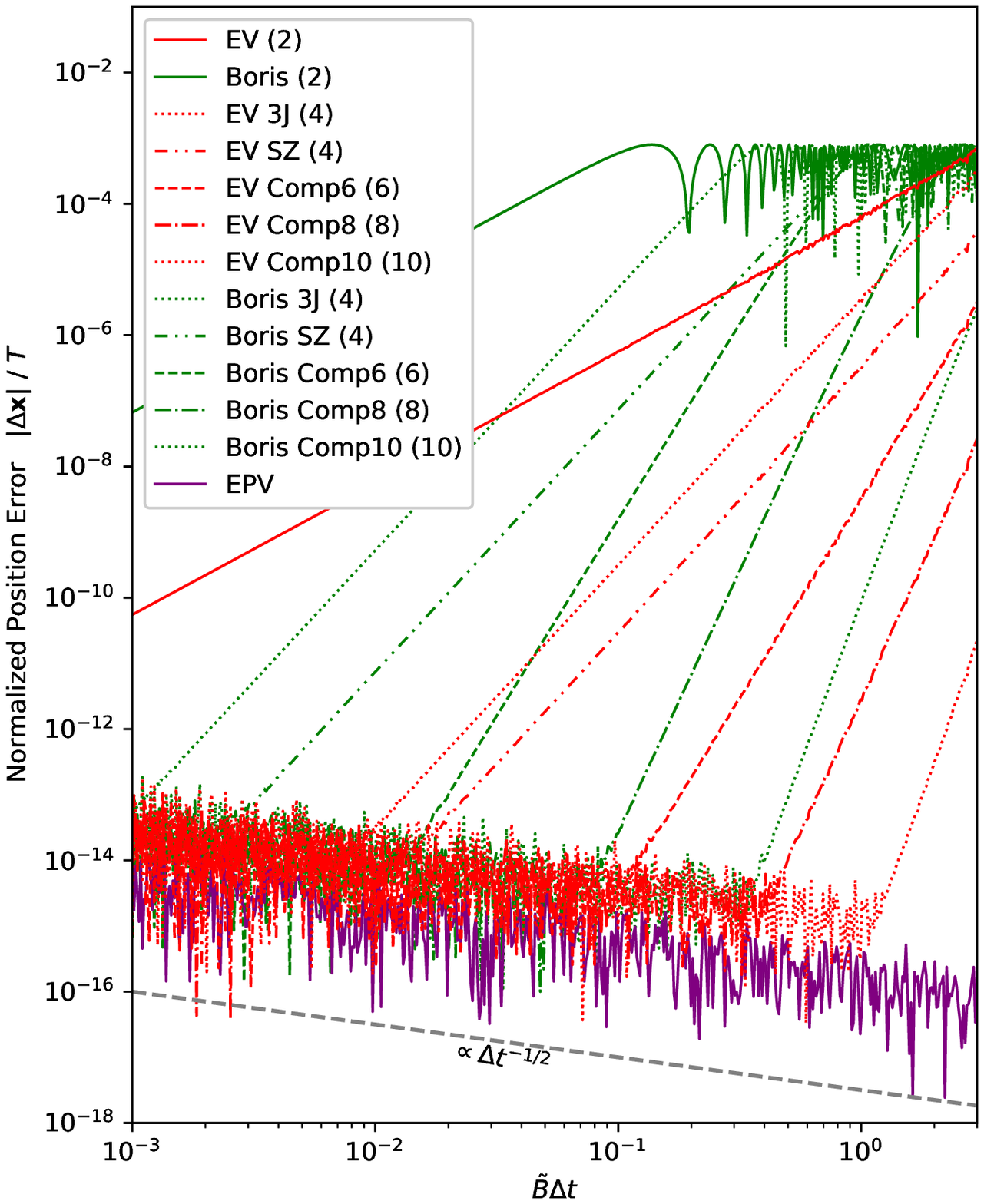}
\caption{
Similar to Fig.~\ref{fig:ExB_pos_err}.
Results of higher-order methods obtained by the composition method are shown
in conjunction with results of the exact velocity integrator
and Boris integrator as references.
The numbers in parentheses in the legends represent the order of accuracy of the integrators.
The gray dashed line in the bottom represents the $\Delta t^{-1/2}$ dependence.
}
\label{fig:ExB_pos_err_higher_order}
\end{figure}

Similar to Fig.~\ref{fig:ExB_pos_err},
Figure~\ref{fig:ExB_pos_err_higher_order} shows the normalized position errors of various higher-order methods
obtained in Subsection \ref{subsec:composition_method}
by applying the composition method
with the exact velocity integrator and Boris integrator as the basic methods.
The triple jump method and Suzuki's fractal composition method with the Boris integrator as the basic method
display fourth-order overall accuracy.
Moreover, these result in
smaller errors than the exact velocity integrator without the composition method
(which displays second-order overall accuracy)
in the region where $\Delta t$ is small.
However, the error can be larger than this where $\Delta t$ is large.
The results obtained using the symmetric composition method of the sixth-order (Comp6),
eighth-order (Comp8), and tenth-order (Comp10) are also shown.
The higher the order, the higher is the accuracy.
However, for the same order,
a significantly higher accuracy is achieved by applying the composition method to the exact velocity integrator
than by applying it to the Boris integrator, as the basic method
(e.g., higher by four orders of magnitude for the triple jump method and
by at least six orders of magnitude for the Comp6 and Comp8 methods)
.
In particular,
it can be observed that the exact velocity integrator with Comp10
calculates with almost the maximum precision
(except for the accumulation of rounding errors) for $\tilde{B}\Delta t < 1$.

\begin{figure}
\includegraphics[width=16cm,clip]{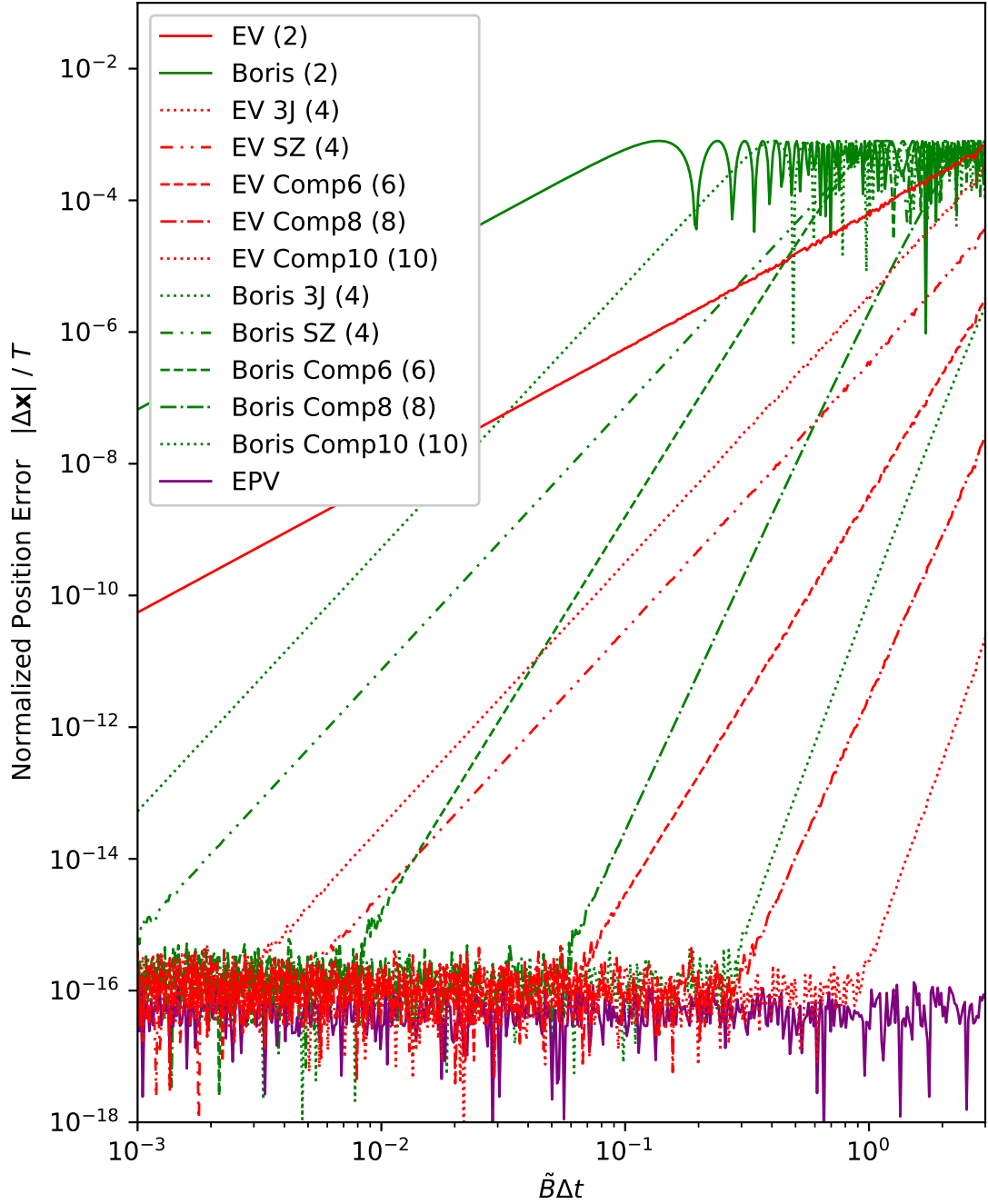}
\caption{
Similar to Fig.~\ref{fig:ExB_pos_err_higher_order}.
Results of higher-order methods using the compensated summation
are shown.
}
\label{fig:ExB_pos_err_higher_order_CS}
\end{figure}

For the high-precision integrators,
the lower limit of the error is determined by the accumulation of rounding errors.
As shown in Fig.~\ref{fig:ExB_pos_err_higher_order},
the smaller the time-step $\Delta t$,
the larger is the accumulation error.
This is because the number of steps increases with a decrease in $\Delta t$
for an equal simulation time $T$.
The dependence of the accumulated rounding errors
on $\Delta t$ appears to be $\propto \Delta t^{-1/2}$.
This is in agreement with their probabilistic explanation 
(for a fixed calculation time $t$)
in Section VIII.~5 of Ref.~\onlinecite{Hairer2006}.
This $\Delta t^{-1/2}$ dependence is also illustrated
by the gray dashed line at the bottom of the figure.
As mentioned in Subsection \ref{subsec:compensated_summation},
the rounding error can be improved using the compensated summation
at the expense of calculation cost.
Figure~\ref{fig:ExB_pos_err_higher_order_CS} shows the results calculated
using this technique with the same simulation parameters.
It can be observed that
the accumulation error in Fig.~\ref{fig:ExB_pos_err_higher_order}
is removed almost completely in Fig.~\ref{fig:ExB_pos_err_higher_order_CS}.
For example,
the error at $\tilde{B}\Delta t = 10^{-3}$
is approximately $10^{-14}$ in Fig.~\ref{fig:ExB_pos_err_higher_order},
whereas
it is approximately $10^{-16}$ in Fig.~\ref{fig:ExB_pos_err_higher_order_CS},
which is almost the machine epsilon in this case.

\subsection{Gyro motion test}
To effectively determine the accuracy of the phase,
it would be more suitable
to investigate the pure gyro motion in a constant magnetic field
without electric fields.
The results of such a gyro motion simulation are shown here.
The simulation parameters are
identical to those in the $E \times B$ drift simulation in the previous subsection,
except that the electric field is set to zero.

\begin{figure}
\includegraphics[width=16cm,clip]{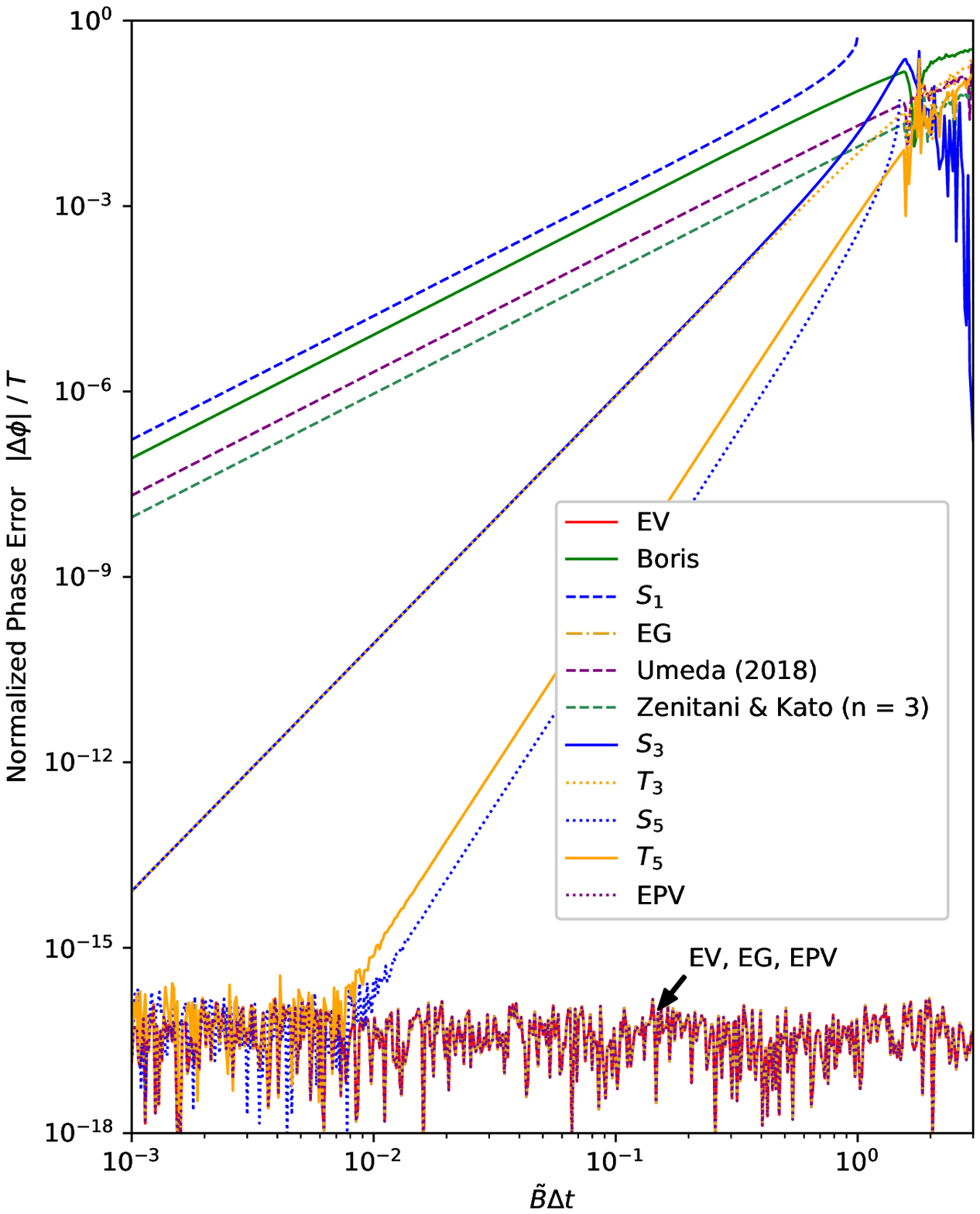}
\caption{
Magnitude of the phase error in velocity per unit time
(which is the deviation of the velocity phase from the exact solution, $|\Delta \phi |$,
divided by the calculation time $T$)
as a function of $\theta = \tilde{B}\Delta t$.
}
\label{fig:B_phase_err}
\end{figure}

Figure \ref{fig:B_phase_err} shows
the magnitude of the phase error in velocity per unit time for various methods.
It is the deviation of the velocity phase from the exact solution, $|\Delta \phi |$,
divided by the calculation time $T=2000$.
As anticipated,
the exact velocity integrator, exact gyration integrator,
and exact position-velocity integrator
solve the phase almost exactly.
For the $S_n$- and $T_n$ -methods,
the accuracies corresponding to the order of approximation are obtained.

\subsection{Test in a static, non-uniform electromagnetic field}
\begin{figure}
\includegraphics[width=16cm,clip]{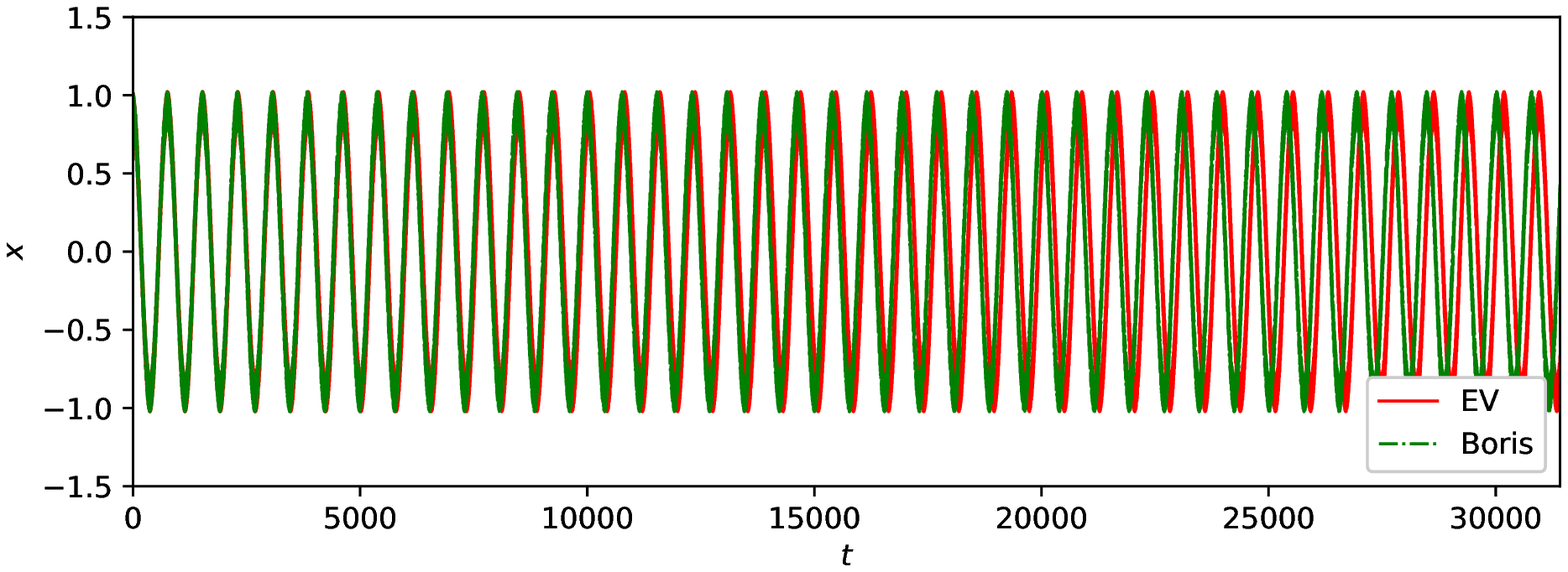}
\caption{
$x$-coordinate of the particle versus time.
Results of the exact velocity integrator (EV)
and Boris integrator are shown.
}
\label{fig:Qin_test_t_x}
\end{figure}

Figure~\ref{fig:Qin_test_t_x} shows the results of the test simulation
in a static, non-uniform electromagnetic field.
The electromagnetic field structure is identical to that in Qin et al. (2013):
\begin{equation}
\bm{B} = (x^2 + y^2)^{1/2} \bm{e}_z,
\qquad
\phi = 10^{-2} (x^2 + y^2)^{-1/2}.
\end{equation}

\begin{figure}
\includegraphics[width=16cm,clip]{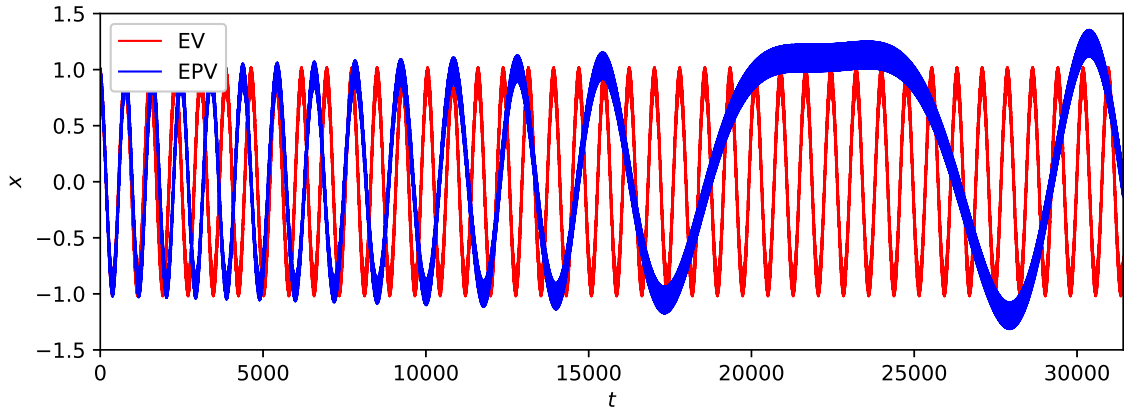}
\caption{
Similar to Fig.~\ref{fig:Qin_test_t_x}.
Results of the exact velocity integrator (EV)
and
exact position-velocity integrator (EPV) are shown.
}
\label{fig:Qin_test_t_x_EPV}
\end{figure}

As can be observed in Fig.~\ref{fig:Qin_test_t_x},
both exact velocity integrator and Boris integrator
maintain the amplitude of the trajectory
even after a long time of integration.
Although not shown in the figure,
this applies to the $S_n$- and $T_n$-methods as well.
This is because these are volume-preserving methods.
Meanwhile, as shown in Fig.~\ref{fig:Qin_test_t_x_EPV},
the trajectory obtained by the exact position-velocity integrator
(for which the volume-preservation condition is not satisfied)
deviates significantly over time.

\subsection{Calculation time}
\begin{figure}
\includegraphics[width=16cm,clip]{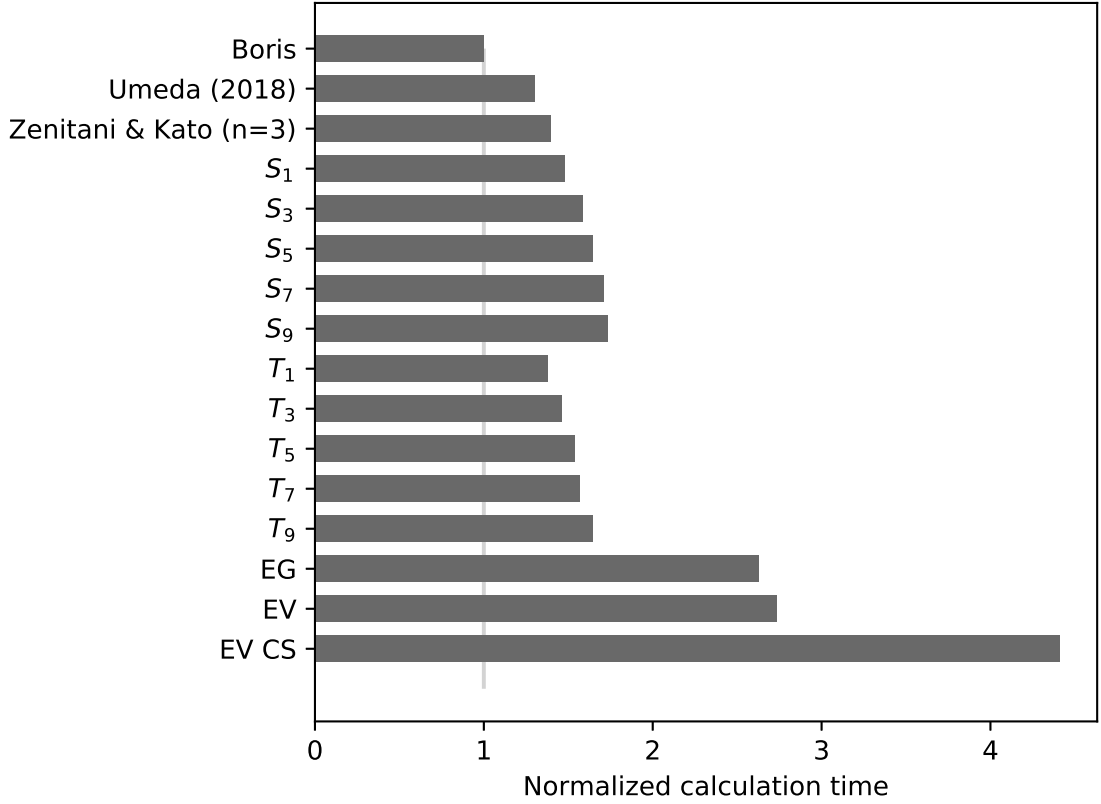}
\caption{
Normalized calculation time of each integrator
in the $E \times B$ drift simulation.
The abbreviations of the integrators are identical to those in the previous figures.
The exact velocity integrator with the compensated summation
is denoted by ``EV CS.''
}
\label{fig:ExB_calc_time}
\end{figure}

Figure~\ref{fig:ExB_calc_time} shows
the average calculation time of each integrator
in the $E \times B$ drift simulation normalized by that of the Boris integrator.
The simulation settings are identical to those in Subsection \ref{subsec:ExB_drift_test}.
The simulation program was compiled with the Intel C++ compiler (version 19.1),
and the simulations were performed on an Intel Core i9-9900K processor.
For each integrator, the calculation time is obtained
by averaging the values for 500 simulations for 10000 particles.
The exact velocity and exact gyration integrators, which use the sine function in their calculations,
consumes 2.5 times longer time than the Boris integrator to calculate.
The exact velocity integrator with the compensated summation
consume four times longer time.
Other integrators consume 1.3--1.7 times longer time than the Boris integrator.
(Because these calculation times do not include
the interpolation time of the electromagnetic field at the particle position,
the time ratios with respect to the Boris integrator would be smaller in the actual PIC simulations.)
Among these approximate integrators,
the $T_5$-, $T_7$-, and $T_9$-methods exhibit good trade-off between speed and precision.
These have computational accuracy almost equal to that of the exact velocity integrator
under the general selection of $\Delta t$ (say, $\tilde{B} \Delta t < 0.5$),
while significantly reducing the computational time.
Note that
as mentioned previously,
the calculation result of the $T_1$-method is identical to that of the Boris integrator.
However, the calculation time is longer because the calculation procedure is different.

The calculation times for the higher-order integrators (not shown here)
are approximately proportional to the number of compositions of the basic integrator.
That is,
the calculation times of the triple jump method, Suzuki's fractal composition method,
and the symmetric composition methods of orders 6, 8, and 10 are approximately 3, 5, 7, 15, and 35 times, respectively, longer than
that of the basic integrator.

\section{Concluding remarks}
In this study,
we constructed the exact velocity integrator
for non-relativistic particles
based on the exact flow of the equation of motion of particles
in a constant electromagnetic field
by means of the splitting method.
We also derived approximate integrators and high-precision integrators
from it. 
All these integrators are volume-preserving
and are suitable for long-term integrations,
particularly in PIC simulations.
The results of numerical tests showed that
the exact velocity integrator is significantly more accurate than the existing integrators
that are volume-preserving and has second-order accuracy.
For example,
in the $E \times B$ drift test,
it was shown that
this integrator is more accurate than the Boris integrator
and exact gyration integrator by three and two orders of magnitude, respectively.
If the calculation speed is more important than accuracy,
the approximate integrators (the $S_n$-methods and $T_n$-methods)
may be useful.
In particular,
the $T_5$-, $T_7$-, and $T_9$-methods exhibit good trade-off between speed and precision.
The calculation cost is slightly higher than that of the Boris integrator.
However, the accuracy is improved substantially (by over three orders of magnitude)
to be almost equal to that of the exact velocity integrator in general cases.
It was also shown that the Boris integrator is equivalent to the $T_1$-method.
In contrast,
when the calculation accuracy is more important than the calculation speed,
the high-precision integrators obtained by applying the composition methods 
to the exact velocity integrator can be used.
In this study, we derived these up to tenth-order accuracy.

The exact position-velocity integrator,
which uses the exact solution of position in addition to that of velocity,
does not satisfy the volume-preservation condition
and is not suitable for long-term integrations.
However,
the exact solution of position (\ref{eq:x_solution_3}) can be applied to determine current density
of a particle.
With this solution,
the displacement of the particle during the time $\Delta t$ is given by
\begin{equation}
\Delta \bm{x} = \bm{v}(0)\Delta t
+ f_2 \bm{e}_1
+ f_3 \bm{e}_2
+ \frac{\frac{1}{2}\Delta t^2 - f_2}{\tilde{B}^2} \bm{e}_3
\end{equation}
and
the exact mean velocity during the time interval $\Delta t$, $\bar{\bm{v}}$,
is obtained with this as $\bar{\bm{v}} = \Delta \bm{x}/\Delta t$.
The contribution of this particle to the current density
can be given by $\bm{j} = q \bar{\bm{v}}$,
and it may improve the accuracy of PIC simulations.

We intend to investigate
the construction of a volume-preserving integrator for relativistic cases
in a similar manner in future work.

\begin{acknowledgments}
This work was partially supported by JSPS KAKENHI (Grant Numbers JP17H02877 and JP21K03627).
\end{acknowledgments}

\appendix
\section{Derivation of exact solution of equation of motion in a constant electromagnetic field}
\label{sec:exact_solution_of_eom}

Although the exact solution of Eq.~(\ref{eq:eom2}) in a constant electromagnetic field would be well established,
we derive it here in a form that is effective for numerical calculation.
Eq.~(\ref{eq:eom2}) is split into the following two equations by resolving the velocity into two components parallel and perpendicular to the magnetic field ($\bm{v}_\parallel$ and $\bm{v}_\perp$, respectively):
\begin{equation}
\frac{d\bm{v}_\parallel}{ds} = \tilde{\bm{E}}_\parallel,
\qquad
\frac{d \bm{v}_\perp}{ds} = \tilde{\bm{E}}_\perp + \bm{v}_\perp \times \tilde{\bm{B}}.
\label{eq:eom3}
\end{equation}
The parallel part can be solved as
\begin{equation}
\bm{v}_\parallel(s) = \bm{v}_\parallel(0) + \tilde{\bm{E}}_\parallel s.
\label{eq:v_solution_parallel}
\end{equation}
For the perpendicular part,
if we introduce
\begin{equation}
\bm{q} \equiv \tilde{\bm{E}}_\perp + \bm{v}_\perp \times \tilde{\bm{B}},
\label{eq:q_def}
\end{equation}
the second equation in (\ref{eq:eom3}) can be rewritten as
\begin{equation}
\frac{d\bm{q}}{ds} = \bm{q} \times \tilde{\bm{B}}.
\end{equation}
This equation is identical to that for particle motion
in a constant magnetic field without electric field.
The solution is given by the well-known gyro motion:
\begin{equation}
\bm{q}(s) = \cos\theta \bm{q}(0) + \sin\theta \bm{q}(0) \times \hat{b},
\end{equation}
where $\tilde{B} \equiv |\tilde{\bm{B}}|$ and
\begin{equation}
\theta \equiv \tilde{B}s,
\quad
\hat{b} \equiv \frac{\tilde{\bm{B}}}{\tilde{B}}.
\end{equation}
Using the following relationship,
\begin{equation}
\bm{v}_\perp = \frac{\tilde{\bm{E}}_\perp \times \tilde{\bm{B}}}{\tilde{B}^2} - \frac{\bm{q} \times \tilde{\bm{B}}}{\tilde{B}^2},
\end{equation}
which is obtained by taking the cross-product of Eq.~(\ref{eq:q_def}) with $\tilde{\bm{B}}$,
we obtain the solution of the perpendicular part:
\begin{equation}
\bm{v}_\perp(s) = \bm{v}_\perp(0)
+ \frac{\sin\theta}{\tilde{B}}\bm{q}(0)
+ \frac{1-\cos\theta}{\tilde{B}^2} \bm{q}(0) \times \tilde{\bm{B}}.
\label{eq:v_solution_perp}
\end{equation}
Finally, by combining Eqs.~(\ref{eq:v_solution_parallel}) and (\ref{eq:v_solution_perp}),
we obtain
\begin{equation}
\bm{v}(s) = \bm{v}(0)
+ \frac{\sin\theta}{\tilde{B}}\bm{q}(0)
+ \frac{1 - \cos\theta}{\tilde{B}^2}\bm{q}(0)\times\bm{\tilde{B}} + \tilde{\bm{E}}_\parallel s.
\label{eq:v_solution}
\end{equation}
This is the exact solution of the particle velocity in a constant electromagnetic field.

By defining
\begin{equation}
\bm{a}(s) \equiv \tilde{\bm{E}} + \bm{v}(s) \times \tilde{\bm{B}} = \bm{q}(s) + \tilde{\bm{E}}_\parallel,
\end{equation}
we can rewrite Eq.~(\ref{eq:v_solution}) as
\begin{equation}
\bm{v}(s) = \bm{v}(0)
+ \frac{\sin\theta}{\tilde{B}}\bm{a}(0)
+ \frac{1 - \cos\theta}{\tilde{B}^2}\bm{a}(0)\times\bm{\tilde{B}}
+ \frac{\theta - \sin\theta}{\tilde{B}^3}(\tilde{\bm{E}}\cdot\tilde{\bm{B}})\tilde{\bm{B}}
\label{eq:v_solution_2}
\end{equation}
This equation can be conveniently integrated with respect to $s$ 
to obtain the exact solution of particle position:
\begin{equation}
\bm{x}(s) = \bm{x}(0)
+ \bm{v}(0) s
+ \frac{1 - \cos\theta}{\tilde{B}^2}\bm{a}(0)
+ \frac{\theta - \sin\theta}{\tilde{B}^3}\bm{a}(0)\times\bm{\tilde{B}}\\
+ \frac{\frac{1}{2}\theta^2 -1 + \cos\theta}{\tilde{B}^4}(\tilde{\bm{E}}\cdot\tilde{\bm{B}})\tilde{\bm{B}}
\end{equation}

For convenience,
we introduce the factors
\begin{equation}
f_1 \equiv \frac{\sin\theta}{\tilde{B}},
\qquad
f_2 \equiv \frac{1 - \cos\theta}{\tilde{B}^2},
\qquad
f_3 \equiv \frac{\theta - \sin\theta}{\tilde{B}^3}
\end{equation}
and the ``bases''
\begin{equation}
\bm{e}_1 \equiv \bm{a}(0) = \tilde{\bm{E}} + \bm{v}(0) \times \tilde{\bm{B}},
\qquad
\bm{e}_2 \equiv \bm{e}_1 \times \tilde{\bm{B}},
\qquad
\bm{e}_3 \equiv (\tilde{\bm{E}}\cdot \tilde{\bm{B}}) \tilde{\bm{B}}.
\end{equation}
With these,
we finally obtain the following expressions for the exact solutions
of velocity and position:
\begin{equation}
\bm{v}(s) = \bm{v}(0)
+ f_1 \bm{e}_1
+ f_2 \bm{e}_2
+ f_3 \bm{e}_3
\end{equation}
\begin{equation}
\bm{x}(s) = \bm{x}(0) + \bm{v}(0)s
+ f_2 \bm{e}_1
+ f_3 \bm{e}_2
+ \frac{\frac{1}{2}s^2 - f_2}{\tilde{B}^2} \bm{e}_3.
\end{equation}

\section{Factors for symmetric composition methods}
\label{sec:factor_symm_comp}

The values of the factors $\gamma_i$ for the symmetric composition method
of orders 6, 8, and 10 are given in Ref.~\onlinecite{Hairer2006}
(also refer the original papers \cite{Yoshida1990, Suzuki1994, McLachlan1995, Sofroniou2005}).
In this study,
we use the values in the following tables
within the precision of the double-precision floating-point number.

\begin{table}[htb]
\caption{Factors for symmetric composition method of order 6}
\centering
\begin{tabular}{rr}
\hline
$\gamma_1 = \gamma_7 =$ & 0.78451361047755726381949763\\
$\gamma_2 = \gamma_6 =$ & 0.23557321335935813368479318\\
$\gamma_3 = \gamma_5 =$ & -1.17767998417887100694641568\\
$\gamma_4 =$ & 1.31518632068391121888424973\\
\hline
\end{tabular}
\end{table}

\begin{table}[htb]
\caption{Factors for symmetric composition method of order 8}
\centering
\begin{tabular}{rr}
\hline
$\gamma_1 = \gamma_{15} =$ & 0.74167036435061295344822780\\
$\gamma_2 = \gamma_{14} =$ & -0.40910082580003159399730010\\
$\gamma_3 = \gamma_{13} =$ & 0.19075471029623837995387626\\
$\gamma_4 = \gamma_{12} = $ & -0.57386247111608226665638773\\
$\gamma_5 = \gamma_{11} =$ & 0.29906418130365592384446354\\
$\gamma_6 = \gamma_{10} =$ & 0.33462491824529818378495798\\
$\gamma_7 = \gamma_9 =$ & 0.31529309239676659663205666\\
$\gamma_8 =$ & -0.79688793935291635401978884\\
\hline
\end{tabular}
\end{table}

\begin{table}[htb]
\caption{Factors for symmetric composition method of order 10}
\centering
\begin{tabular}{rr}
\hline
$\gamma_1 = \gamma_{35} =$ & 0.07879572252168641926390768\\
$\gamma_2 = \gamma_{34} =$ & 0.31309610341510852776481247\\
$\gamma_3 = \gamma_{33} =$ & 0.02791838323507806610952027\\
$\gamma_4 = \gamma_{32} =$ & -0.22959284159390709415121340\\
$\gamma_5 = \gamma_{31} =$ & 0.13096206107716486317465686\\

$\gamma_6 = \gamma_{30} =$ & -0.26973340565451071434460973\\
$\gamma_7 = \gamma_{29} =$ & 0.07497334315589143566613711\\
$\gamma_8 = \gamma_{28} =$ & 0.11199342399981020488957508\\
$\gamma_9 = \gamma_{27} =$ & 0.36613344954622675119314812\\
$\gamma_{10} = \gamma_{26} =$ & -0.39910563013603589787862981\\

$\gamma_{11} = \gamma_{25} =$ & 0.10308739852747107731580277\\
$\gamma_{12} = \gamma_{24} =$ & 0.41143087395589023782070412\\
$\gamma_{13} = \gamma_{23} =$ & -0.00486636058313526176219566\\
$\gamma_{14} = \gamma_{22} =$ & -0.39203335370863990644808194\\
$\gamma_{15} = \gamma_{21} =$ & 0.05194250296244964703718290\\

$\gamma_{16} = \gamma_{20} =$ & 0.05066509075992449633587434\\
$\gamma_{17} = \gamma_{19} =$ & 0.04967437063972987905456880\\
$\gamma_{18} =$ & 0.04931773575959453791768001\\
\hline
\end{tabular}
\end{table}


\bibliography{references}

\end{document}